# Effect of Interface Style in Peer Review Comments for UML Designs

Scott A. Turner, Manuel A. Pérez-Quiñones, Stephen H. Edwards
Computer Science Department
Virginia Polytechnic Institute and State University
Blacksburg, VA 24061
scturner@vt.edu, {perez, edwards}@cs.vt.edu

**ABSTRACT**
This paper presents our evaluation of using a Tablet-PC to provide peer-review comments in the first year Computer Science course. Our exploration consisted of an evaluation of how students write comments on other students' assignments using three different methods: pen and paper, a Tablet-PC, and a desktop computer. Our ultimate goal is to explore the effect that interface style (Tablet vs. Desktop) has on the quality and quantity of the comments provided.

**Category**: H.1-Models and Principles, User/Machine Systems:[Human Factors]; H.4-Information Systems Applications; Miscellaneous; H.5-INFORMATION INTERFACES AND PRESENTATION (e.g., HCI), User Interfaces

**Terms**: Human Factors, Design

1. INTRODUCTION

The Tablet-PC represents the latest generation of personal computers. Its form-factor is like a laptop computer but uses a pen and a touch sensitive display to allow interaction in a pen-driven style. Microsoft has updated its Windows operating system, and created new applications to make use of the pen in new and creative ways [9]. Classroom use is a good application domain for tablets, particularly when coupled with wireless connectivity. The Tablet-PC has the potential of impacting education in the following ways. First, pen-based computers can be used in creative ways in the classroom to enhance the teaching and learning environment [3], similar to what has been done with PDAs [6]. Second, pen-based computers can be used in the classroom for note-taking [7] and sharing these notes with others [4]. Finally, Tablet-PCs are smaller than desktops with monitors, and thus can be used in a traditional lecture-based classroom in ways that complement lectures [1, 2].

We are interested in using Tablet-PCs to provide peer review for UML designs. Peer-review has been found to be a good way to help students learn from each other and to increase their understanding of coding and design issues by exposing them to alternative designs, and giving them more experience in reading other students' work [5].

As a precursor to the introduction of Tablet-PCs in the classroom to help in Object-Oriented Designs, we decided to explore how students used the pen in the Tablet-PC to provide written comments in a UML design. In particular, we were interested to see what general form the handwriting took when students wrote UML design feedback using three configurations: paper and pen, a Tablet-PC using a pen, and a desktop computer.

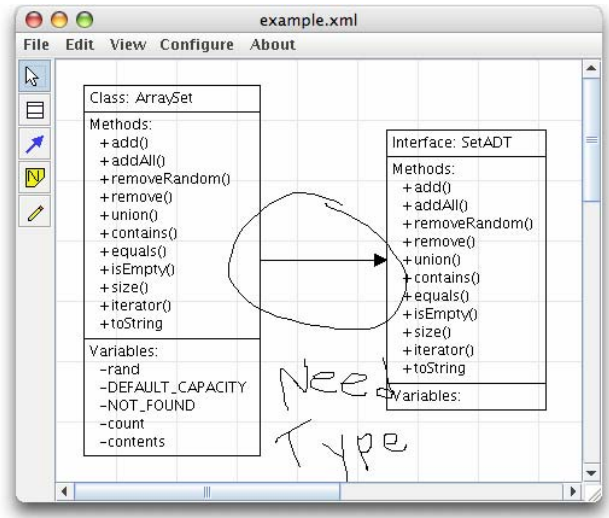

*Figure 1: minimUML Tool*

We use minimUML [8] (see figure 1), a Universal Modeling Language (UML) tool that we have built to support the teaching of object-oriented design in the first year of a Computer Science curriculum[1]. MinimUML is a simple design tool that implements a very small subset of UML. It allows the creation and modification of class objects and the connections between those objects. Annotations are allowed in the form of "comment boxes" which resemble Post-It notes; we call these "sticky" notes. The user can place these notes anywhere in the design and type a text comment inside of the note. The notes can be collapsed into a small icon.

The tool also supports the use of free-form drawings. When the tool is used with a mouse, the free-form drawings are best used to circle or to draw an arrow to point to areas of the design. But, when used with a pen, the free-form can be used as a form of digital ink where the reviewer can write comments directly in the UML design. Both the notes and the free-form drawings can be selected and moved to anywhere in the design. Figure 1 shows an example of comments in minimUML.

The remainder of this document shows the results of a study to observe how students use the pen in a Tablet-PC to provide feedback on an object-oriented design created by other students. In particular we were interested to see how the three different platform configurations produced different types of comments.

2. OUR STUDY

To compare the effect of the use of a pen-based interface with other methods, we conducted a study where we asked students to review three object-oriented UML designs. The review was conducted based on a series of design guidelines given to the students. Using these design guidelines, they were asked to provide feedback by identifying parts of the UML design that were positive (i.e. a good use of a particular design guideline) or negative. Aside from being asked to address the guidelines, the students were allowed to review in a manner that they saw fit.

We had a total of 29 participants, 28 were Computer Science students and 1 was a Computer Engineer at Virginia Tech; 6 were Seniors, 3 Juniors, 19 Graduate students, and 1 did not specify a major. The study was conducted in the Fall of 2004. Participants came to the testing area, signed the consent form, were given instructions, and then proceeded to do three UML design reviews. The total session lasted about 1.5 hours. The study was approved by the Institutional Review Board at Virginia Tech.

---

[1] minimUML is available at http://perez.cs.vt.edu/minimUML/

Each student was asked to review one diagram with a tablet PC the pen input, one with a desktop PC with a mouse, and one with pen and paper. We divided the participants in three groups and each group used the platforms in a different order. All the participants were given the same three UML diagrams but the order in which they received the diagrams was rotated.

## 3. FINDINGS

For this study, our goal was to observe the effect of the choice of platform had on the comments provided by the participants. The nature of the study was mostly observational. The results reported here, therefore, are mostly descriptive in nature.

### 3.1 Number of Comments

The first observation we make is that the number of comments provided varied considerably across the participants and across the platforms. Every platform had at least one student that used it more than the other two and there was a great variety in the relative usage. In some, all three platforms were very close, in others, there was a very clear favorite or favorites. Overall, the pen and paper condition produced significantly ($p < 0.01$) more comments than the other two conditions (with a total of 256). The tablet PC did not have significantly more comments (176) than the desktop (163). Figure 2 shows the number of comments for each platform by each student.

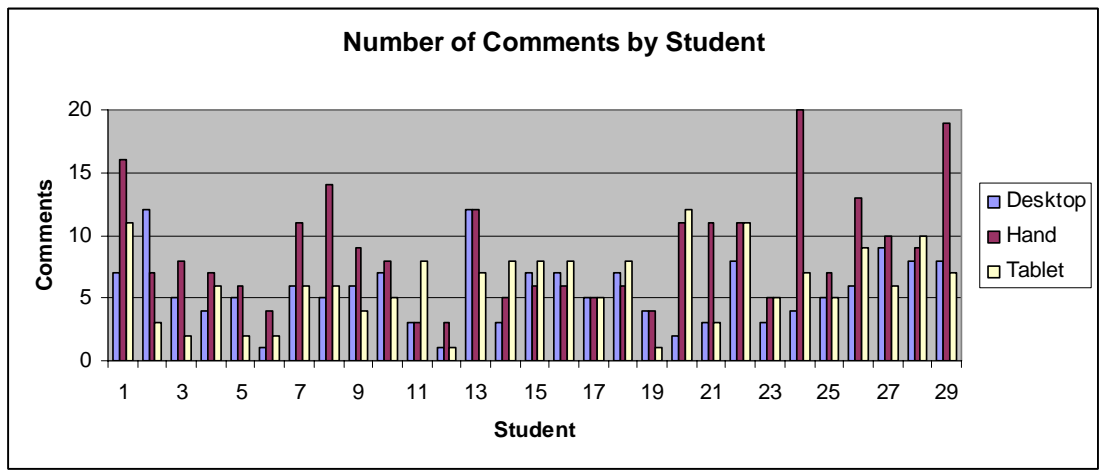

*Figure 2: Average blocks and comments for all participants in all platforms (left) and average comments per participants (right)*

### 3.2 Structure of Comments

Not only did the quantity of the comments vary across platforms, so did the structure of how the comments were provided. This section describes some observations on the structure of the comments.

The comments provided by the students included some or all of the following characteristics.

### 3.2.1 Detailed Editing

Some comments included very detailed editing instructions. In some cases, the reviews included crossed-out words, or even crossed-out letters within names. This was observed mostly with the pen and paper group (see Figures 3 and 4 for examples), and it was also found a few times in the tablet condition.

Figure 3 shows the level of detail of the markup in the paper version. Note how the reviewer crossed out lines of code (bottom right), and at times even crossed out just a few letters (top left). Figure 4 shows similar edits on the tablet.

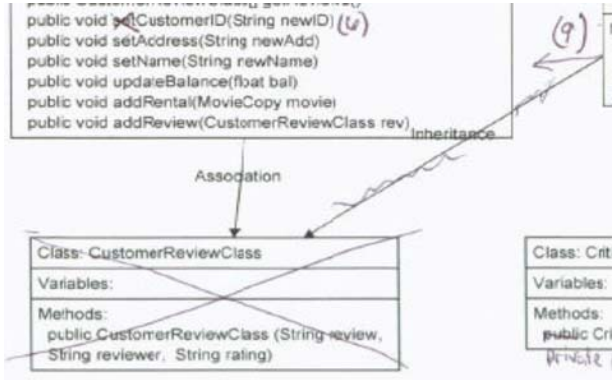 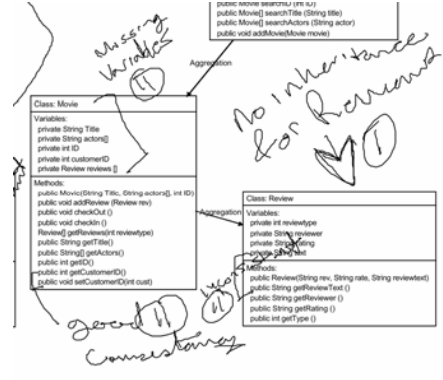

*Figure 3: Detailed editing comments on paper*   *Figure 4: Comment on the tablet*

### 3.2.2 Connection of comments to UML Objects

Comments written by the reviewers were connected to the different UML parts by either gestures or by collocation. Gestures, such as arrows and circles, were used to identify UML components referred to by the comments (which could be across the page). Collocation refers to the placement of the comments in relation to the referent. Simply writing or putting a sticky note on the UML class is sufficient for identification.

The collocation of comments without the use of gestures was typical in the desktop condition, but also appeared in the other two conditions. Gestures were very common in the paper and tablet condition but understandably rare on the desktop. For both the pen and paper and the tablet conditions, it was common to find the use of both collocation and of gestures for a single comment, but it interesting to note that there were more strictly gestural references on the tablet. Figure 5 shows the occurrences of collocation and gestural references as a percent of the total comments for each platform.

### 3.2.3 References

Another way in which the three conditions differed was in how the content of the comments referred to parts of the UML diagrams. We classified as an "explicit" reference a comment that included the name of a class, association or any other element of the design. We classified as an "implicit" reference a comment that used an anaphoric expression. An anaphoric expression is one that requires the context of where the comments were placed to interpret it. For example, a comment could include references to "this class" without naming the class. To interpret "this class," the comment could be positioned next to a class. This type we call "collocation" reference. Another interpretation of "this class" would be with the use of a gesture, such as an arrow pointing from the note to a class. This we call "gestural" reference.

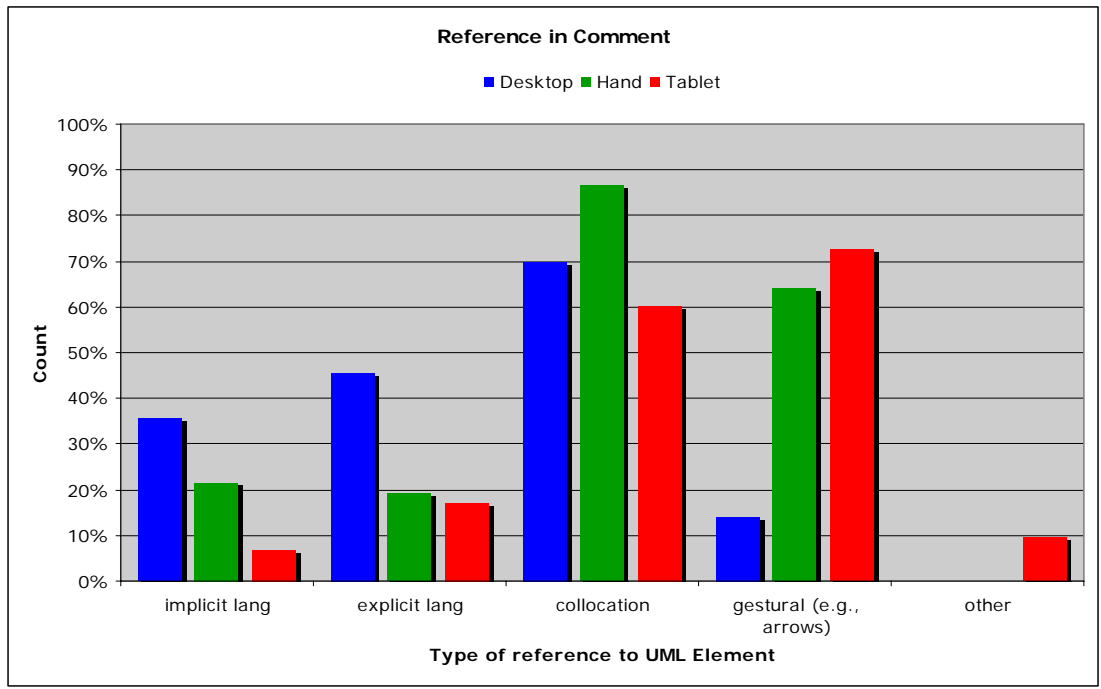

*Figure 5: Distribution of the use of language and how comments were related to the UML designs*

Figure 5 shows the percentages for each of the conditions. Note that the desktop used the most explicit and implicit types of comments, and that paper and tablet tended to use only collocation or gestures as references. We found that most of these differences were significant. In the desktop condition, significantly more implicit language was used than in paper condition (p < 0.005). Paper, in turn, produced more than the tablet (p < 0.001). For the use of explicit references, the desktop, once again, had more (p < 0.001). There was no difference between paper and tablet here.

The use of collocation was more common on paper than the other two conditions (p < 0.001). The tablet used collocation fewer time than the desktop, but not significantly so. Gestural references, not surprisingly, occurred more on the tablet and paper than on the desktop (p < 0.001).

The difficulty in creating gestures on the desktop would account for the higher usage of both implicit and explicit language. This may indicate that the use of gestures makes for simpler comments. It would also appear that, while collocation seems to be an important part of many of the references, it was not enough to be a complete reference in most cases.

### 3.2.4 Content

While examining the contents of the comments, we noted if the comment contained a reference to the guideline that the comment was based on (which we asked the participants to do). We also recorded if the content was verbal, or if it was strictly gestural. That is, the comment was just the crossing out a method with no other explanation.

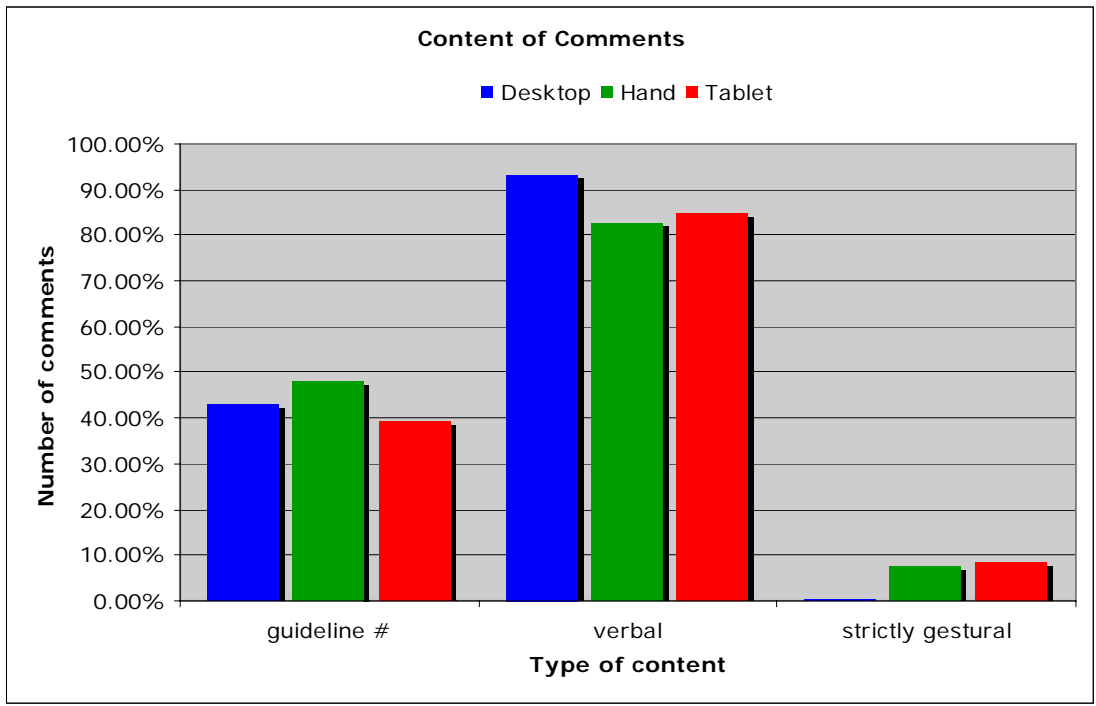

*Figure 6: Distribution of the content of the comments*

The results were hardly surprising. The use of guidelines did not differ significantly between the conditions. The use of verbal comments was more common on the desktop and strictly gestural comments were more common on both the tablet and paper (p < 0.01 for both). It is easy to see that strictly gestural comments are more common on the tablet and paper because it is much easier and more natural to make them on those platforms than on the desktop. That reasoning also accounts for the higher use of verbal comments on the desktop.

3.2.5 Referents

In this study we also examined what each comment referred to. We classified each of the referents as a class, a method or field within a class, an association between classes, or as a general comment for the entire diagram. A small number of comments did not fit into any of these categories and so they were marked as "other."

While there was some variation between the conditions, very little of it was significant. There were no differences in the way the comments referred to classes, methods, fields, or associations. However, there was a significant difference in the number of general comments made on the tablet relative to the other two (p < 0.01). It is possible that this is related to the difference in the overall number of comments. However, one would expect to see a difference with the desktop as well, which there was not. It may also be connected to the amount of detailed editing (mentioned in 3.2.1) found on the reviewed diagrams. The number of referents classified as "other" was not significant.

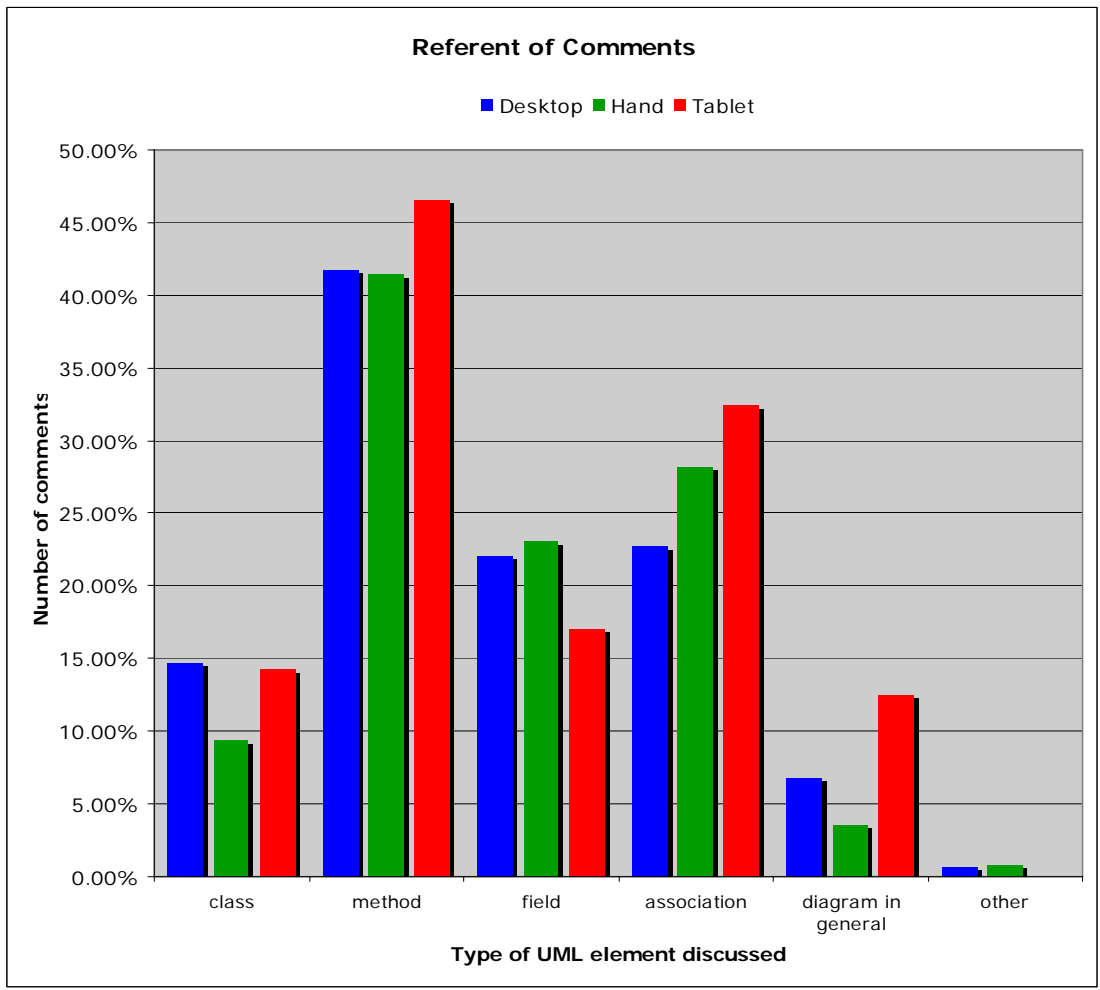

*Figure 7: Distribution of the referents used in the comments*

It is important to note that while there were differences in how the commenting was done, it did not seem to effect the level of detail at which it was done. None of the conditions showed a particular tendency to have comments that were mostly high level or mostly low level. While general comments were more common with the tablet, there were plenty that focused on the method and field level.

4. CONCLUSIONS AND FUTURE WORK

This study allowed us to explore the impact that platform choice had on the structure of comments provided in a peer-review context. For the educational community, we believe that it is important that we consider the human-factors of the technologies that we use in the classroom. Often times, we switch technology without considering that the switch itself makes a significant change on how we and our students do our work.

In general, we found that the most natural medium for providing comments was still the pen and paper. However, that medium also invited more editing-style comments, which were not necessarily appropriate for the task at hand. While the use of editing-style comments was neither encouraged nor discouraged during the task, in a real classroom exercise, it may be beneficial to focus the students on higher level tasks. The Tablet-PC allowed the participants to provide comments that resembled the paper format, while avoiding the detailed level of comments (editing comments) seen in the paper format. The Desktop format was the most restrictive of all (e.g., gestures were very restricted). Furthermore, the comments were almost always provided in stickies and rarely as free-form annotations.

## 5. ACKNOWLEDGMENTS

We would like to thank Tom McMail and Jane Prey at Microsoft University Relations for their support of this work, and Mary Pinney and Alyssa Sams for coding some of the data. Also, we want to acknowledge Jaime García who has been doing some of the development work on minimUML.